\documentclass[12pt,a4paper]{article}
\usepackage[utf8]{inputenc}
\usepackage[T2A]{fontenc}
\usepackage[english]{babel}
\usepackage{amsmath}
\usepackage{amssymb}

\begin{document}
\title{Electromagnetic modulation of monochromatic neutrino beams}
\author{A. L. Barabanov${}^{1,2}$\thanks{\emph{e-mail:} a.l.barabanov@yandex.ru}\,\,, O. A. Titov$^{1}$\thanks{\emph{e-mail:} titov-o@mail.ru}\\
${}^1$ \textit{NRC "Kurchatov Institute"\,, 123182, Moscow, Russia}\\
${}^2$ \textit{Moscow Institute of Physics and Technology, 141700, Dolgoprudny,}\\ \textit{Moscow Region, Russia}}
\date{}
\maketitle

\begin{abstract}

A possibility to produce a modulated monochromatic neutrino beam is discussed. Monochromatic neutrinos can be obtained in electron capture by nuclei of atoms or ions, in particular, by nuclei of hydrogen-like ions. It is shown that monochromatic neutrino beam from such hydrogen-like ions with nuclei of non-zero spin can be modulated because of different probabilities of electron capture from hyperfine states. Modulation arises by means of inducing of electromagnetic transitions between the hyperfine states. Requirements for the hydrogen-like ions with necessary properties are discussed. A list of the appropriate nuclei for such ions is presented.

\end{abstract}

\section{Introduction}

''Clean'' neutrino beams (with neutrinos of single flavour) with known intensity and energy spectrum are of great interest for studying the neutrino properties and the details of neutrino-involving processes. A comprehensive list of applications for such beams is given, e.g., in \cite{Espinoza_2012}. These applications include measurements of neutrino oscillation parameters (in particular, for oscillations to sterile states), search for neutrino magnetic moment, refinement of the weak interaction constants values, study of coherent neutrino scattering off nuclei, investigation of elastic and inelastic neutrino-nucleon and neutrino-nucleus scattering processes (including astrophysical problems).

In \cite{Zucchelli_2002} it was proposed to use accelerated $\beta^+$- and $\beta^-$-\,decaying nuclei (or radioactive ions), held in a storage ring, as a source for such neutrino beams ($\beta$-beams). The neutrinos are emitted isotropically in the rest frame of the nuclei. However, in the laboratory frame, half of the neutrinos are emitted in the forward direction within a cone with the opening angle $\theta\simeq 1/\gamma$. Therefore, if the nuclei are accelerated to $\gamma\gg 1$, the neutrino beam is collimated. The neutrinos, which are emitted forwardly with the energy $E^0_{\nu}$ in the rest frame of the nuclei, have  much larger energy $E_{\nu}\simeq 2\gamma E^0_{\nu}\gg E^0_{\nu}$ in the laboratory frame. So, one can control the neutrino spectrum by varying the value of $\gamma$.

A useful modification has been advanced in \cite{Sato_2005,Bernabeu_2005}: one can use the ions with electron-capturing nuclei as a neutrino source. The advantage of the electron capture (EC) is that in the ion rest frame the neutrino energy $E_{\nu}^0$ is completely definite. Thus, in the laboratory frame at $\gamma\gg 1$, one can obtain a high-energy neutrino beam, which is collimated and monochromatic (EC-beam).

In this paper, we discuss another possible improvement: EC-beam could be modulated, or, in other words, transformed into a sequence of neutrino pulses. Such a beam, in principle, may be used to transmit information (regarding neutrino communications see \cite{Stancil_2012} and references therein). However, using modulation to increase the efficiency of neutrino detection seems more realistic. A modulated beam causes a modulated signal in the detector which can be separated from constant or randomly fluctuating background noise (the similar advantage is pointed out in \cite{Ginzburg_2015} for neutrinos produced by interaction of a modulated high-energy electron beam with a target).

Our proposal is closely related to the results, obtained at GSI (Darmstadt) in experiments with highly-charged ions with $\beta^+$-radioactive nuclei. The ESR (Expe\-ri\-men\-tal Storage Ring) setup was used to study EC decay of short-living hydrogen-like (H-like) ions of $^{140}$Pr \cite{Litvinov_2007,Litvinov_2008}, $^{142}$Pm \cite{Litvinov_2008,Winckler_2009,Kienle_2013} and $^{122}$I \cite{Atanasov_2012}. These nuclei decay with comparable probabilities both through $\beta^+$ channel, emitting a positron $e^+$ and a neutrino $\nu_e$, as well as through EC channel, capturing an electron $e^-$ and emitting a monochromatic neutrino $\nu_e$. The ions were produced in fragmentation reactions without additional acceleration, therefore, they were stored at moderate energies ($\gamma\simeq 1.4$). The decays were detected by observing the disappearance of mother nuclei and the appearance of daughter nuclei in the ESR; the emitted neutrinos were not observed directly.

One of the goals of these experiments was to study a dependence of the EC decay rate upon the number of electrons in the ion; not only H-like ions were investigated, but also He-like ions. At first glance, a paradoxal result was obtained: for H-like ions $^{140}$Pr, $^{142}$Pm and $^{122}$I the EC decay constant $\lambda_{EC}$ (decay probability per unit time) turned out to be larger than it is for the corresponding He-like ions! However, an explanation was given right away in \cite{Litvinov_2007}. The effect is due to the hyperfine splitting of the H-like ion ground state (in all cases the nuclear spin $I\ne 0$). A more detailed analysis is presented in \cite{Patyk_2008,Ivanov_2008,Siegien_2011}.

It is worthwhile to note that the capture of an electron $e^-$ by the nucleus in a H-like ion (with emission of an electron neutrino $\nu_e$) is similar to the muon capture by the nucleus in a muonic atom (with emission of a muonic neutrino $\nu_{\mu}$). In both cases, a lepton of spin 1/2 is captured from the $|1s\rangle$\,-\,state. In more exact terms, if the nucleus has a non-zero spin $I$, then the capture occurs from the hyperfine state with definite total angular momentum $F=I\pm 1/2$. In general, the decay probability highly depends on the value of $F$. This dependence for muon capture was first found out in 1950s \cite{Bernstein_1958,Primakoff_1959} and it was called ''the hyperfine effect''. This effect was repeatedly observed (see, e.g., \cite{Wiaux_2002} and references therein).

The hyperfine effect is most pronounced in pure Gamow--Teller transitions, i.e.
\begin{equation}
\label{1}
I^{\pi}\to I^{\prime\,\pi},\qquad I^{\prime}=I\pm 1,
\end{equation}
where $I^{\prime}$ is the daughter nucleus spin. In the Gamow--Teller approximation, the total angular momentum consists of the daughter nucleus spin $I^{\prime}$ and the neutrino spin 1/2. Therefore, due to the total angular momentum conservation the following condition must be satisfied:
\begin{equation}
\label{2}
F = I' \pm 1/2.
\end{equation}
This means that the mother nucleus can capture $e^-$ or $\mu^{-}$ from only one of the hyperfine states: if $I'=I-1$, the capture occurs from the state $F = I-1/2$, while in the case of $I'=I+1$ the capture occurs from the state $F = I+1/2$ (see Fig.~\ref{f1}; more detailed comments for this figure are given below).  

\begin{figure}

\begin{center}
\begin{picture}(130,80)

\put(0,0){\line(1,0){130}}
\put(0,35){\line(1,0){130}}
\put(0,70){\line(1,0){130}}

\put(40,0){\line(0,1){80}}
\put(85,0){\line(0,1){80}}

\put(2,75){Type of H-like ion}
\put(57,75){$\mu>0$}
\put(102,75){$\mu<0$}
\put(17,50){A}
\put(17,15){F}

\put(43,62){\line(1,0){15}}
\put(43,57){\line(1,0){15}}
\put(43,42){\line(1,0){15}}

\put(60,61){$F=I+1/2$}
\put(60,56){$F=I-1/2$}
\put(60,41){$I^{\prime}=I-1$}

\put(48,57){\vector(0,-1){15}}
\put(53,62){\vector(0,-1){20}}
\put(51.6,52){$\times$}

\put(88,62){\line(1,0){15}}
\put(88,57){\line(1,0){15}}
\put(88,42){\line(1,0){15}}

\put(105,61){$F=I-1/2$}
\put(105,56){$F=I+1/2$}
\put(105,41){$I^{\prime}=I+1$}

\put(93,57){\vector(0,-1){15}}
\put(98,62){\vector(0,-1){20}}
\put(96.6,52){$\times$}

\put(43,27){\line(1,0){15}}
\put(43,22){\line(1,0){15}}
\put(43,7){\line(1,0){15}}

\put(60,26){$F=I+1/2$}
\put(60,21){$F=I-1/2$}
\put(60,6){$I^{\prime}=I+1$}

\put(48,22){\vector(0,-1){15}}
\put(53,27){\vector(0,-1){20}}
\put(46.6,14.5){$\times$}

\put(88,27){\line(1,0){15}}
\put(88,22){\line(1,0){15}}
\put(88,7){\line(1,0){15}}

\put(105,26){$F=I-1/2$}
\put(105,21){$F=I+1/2$}
\put(105,6){$I^{\prime}=I-1$}

\put(93,22){\vector(0,-1){15}}
\put(98,27){\vector(0,-1){20}}
\put(91.6,14.5){$\times$}

\end{picture}
\end{center}

\caption{\label{f1} Schematic view of allowed and forbidden (marked with $\times$ sign) transitions of pure Gamow--Teller type for electron capture in H-like ions, depending on the sign of magnetic moment $\mu$ of the mother nucleus and on the relation between the spins $I$ and $I'$ of the initial and final nuclear states. The ground state of an initial H-like ion is splitted to the states with total angular momenta $F=I\pm 1/2$ due to the hyperfine interaction. The ion is either of A-type or F-type, depending on whether the electron capture is allowed or forbidden for the lower hyperfine state --- see Section~\ref{HFS} below.}
\end{figure}

In the H-like ions of $^{140}$Pr, $^{142}$Pm and $^{122}$I, the hyperfine effect does take place. Indeed, these nuclei are of spin and parity $1^+$, and they decay into $0^+$-states of the daughter nuclei. Thus, the electron capture in the H-like ion is allowed only from the level $F=1/2$ (the decay from the level $F=3/2$ is forbidden, or, strictly speaking, it is highly suppressed). For the ions specified, the state $F=1/2$ is the ground state, because it lies below the state $F=3/2$.

It is clear that, inducing transitions between the states with ''allowed'' and ''forbidden'' $F$ values, one can impact the capture rate, and, therefore, modulate the corresponding neutrino emission. The idea to use electromagnetic radiation to induce transitions between hyperfine states and change the capture probability was first stated for muonic atoms in \cite{Batkin_1977}. In 1990s, the hyperfine effect for the H-like ions was rediscovered in \cite{Folan_1995} (evidently, due to the prospects for experimental studies of H-like ions with heavy $\beta^+$-radioactive nuclei). The authors of \cite{Folan_1995} also noted the possibility to impact the EC rate by using electromagnetic radiation.

The purpose of this paper is to select the appropriate $\beta^+$-radioactive nuclei (in fact, H-like ions with such nuclei), that could be used as sources of modulated monochromatic neutrino beams. In section~\ref{Isotopes} we state the requirements for the isotopes and give a list of the most suitable ones. Section~\ref{HFS} deals with the hyperfine structure of H-like ions and electromagnetic transitions between the hyperfine states. In section~\ref{Results}, the main properties of the H-like ions with appropriate nuclei are given, and their applicability to produce the modulated monochromatic neutrino beam is discussed. The results are briefly summarized in Section~\ref{Conclusion}.

\section{Possible source isotopes for modulated beams \label{Isotopes}}

As it was stated above, H-like ions with electron-capturing nuclei, trapped in a storage ring, are considered as a source of monochromatic and modulated neutrino beam. The nuclei must have non-zero spin in order for the ground state of H-like ion be splitted due to the hyperfine interaction. Beam modulation is possible, if the EC decay is allowed for one of the hyperfine states and forbidden for the other one.

This situation takes place for the Gamow--Teller transitions (\ref{1}), because the condition (\ref{2}) is satisfied only by one of the hyperfine states. So, the $\beta^+$-radioactive mother nucleus of the H-like ion must be of spin $I\ne 0$, the daughter nucleus spin has to differ from $I$ by 1, and the parities of the initial and final states have to be the same.

The neutrino beam will be entirely monochromatic, provided, first, $\beta^+$ decay (with emission of $e^+$ and $\nu_e$) of the mother nucleus is highly suppressed or not occur at all, and, second, the EC decay (with emission of $\nu_e$) goes with a nuclear transition to only one state of the daughter nucleus. This state can be either the ground state or an excited one. To be more realistic, we take the deviation of 1-2 \% from full monochromaticy of neutrinos, and thus we include into consideration the transitions to nuclear final states with branching of 99-98 \% of the total decay probability.

The EC transition from the ground state of the mother nucleus $^A_ZX$ into a state of the daughter nucleus $\lefteqn{{}_{Z-1}}\,\,\,\,\,{}^AX^{\prime}$ with excitation energy $E^{\,\prime}$ ($E^{\,\prime}=0$ for the ground state) is accompanied by the energy release
\begin{equation}
\label{3}
Q_{EC}=\Delta(A,Z)-\Delta(A,Z-1)-E^{\,\prime},
\end{equation}
where $\Delta(A,Z)$ is the atomic mass excess. The $\beta^+$ transition between the same states gives the energy release
\begin{equation}
\label{4}
Q_{\beta^+}=\Delta(A,Z)-\Delta(A,Z-1)-E^{\,\prime}-2m_ec^2\equiv Q_{EC}-2m_ec^2.
\end{equation}
Thus, we should select the transitions with $Q_{EC}$ value less or just slightly more than the doubled electron rest energy $2m_ec^2$ (in these cases, $\beta^+$ decay is impossible or strongly suppressed).

Note that, since the required energy release $Q_{EC}$ is relatively small, all the nuclides of our interest are close to the nuclear stability region. The characteristics of such nuclei are rather well-known (we use data from \cite{IAEA}).

There are also constraints on half-lives of short-living mother nuclei. They should not decay too rapidly, because some time (near 2 seconds) is needed to trap H-like ions into the storage ring and then to cool them down (see, e.g., \cite{Litvinov_2007,Winckler_2009}). On the other hand, the lifetime should not be too long, because this will result in low decay rate and small intensity of neutrino emission. In practice, the nuclei, decaying only through EC, have lifetimes from approximately a minute and above. Therefore, one needs to set only the upper limit on half-lives. We have set a rather high upper limit
\begin{equation}
\label{5}
T_{1/2} < 10^6~\mbox{s}\simeq 11.6~\mbox{d}.
\end{equation}

Table \ref{t1} shows the nuclear transitions that satisfy the requirements stated above. Entirely monochromatic neutrinos could be produced in three transitions with small $Q_{EC}<2m_ec^2$ between the ground states of mother and daughter nuclei: ${}^{71}_{32}{\rm Ge}\to {}^{71}_{31}{\rm Ga}$, ${}^{131}_{\,\,\,55}{\rm Cs}\to {}^{131}_{\,\,\,54}{\rm Xe}$ and ${}^{165}_{\,\,\,68}{\rm Er}\to {}^{165}_{\,\,\,67}{\rm Ho}$. In all three cases, the final nuclei are stable.

There are two more transitions between the ground states of mother and daughter nuclei with higher $Q_{EC}$ value, ${}^{135}_{\,\,\,57}{\rm La}\to {}^{135}_{\,\,\,56}{\rm Ba}$ and ${}^{163}_{\,\,\,68}{\rm Er}\to {}^{163}_{\,\,\,67}{\rm Ho}$. These transitions account for $98.1$\% and $99.9$\% of the total decay rate, respectively; the remaining fraction is associated with the EC transitions to excited states of the daughter nucleus. Note that one of the specified daughter nuclei, ${}^{135}_{\,\,\,56}$Ba, is stable, while the second one, ${}^{163}_{\,\,\,67}$Ho, is formally unstable: it decays through the EC channel. However, its half-life is very large: $T_{1/2}=4570$~years (and without electron shell the nucleus is stable).

Finally, there are two transitions with rather high $Q_{EC}$ values, ${}^{107}_{\,\,\,48}{\rm Cd}\to {}^{107}_{\,\,\,47}{\rm Ag}^*$ and ${}^{118m}_{\,\,\,\,\,\,\,\,51}{\rm Sb}\to {}^{118}_{\,\,\,50}{\rm Sn}^*$. In both cases, the transition occur mainly to one excited state of the daughter nucleus with $99.7$\% and $98.3$\% fraction,  respectively. Again, the remainder fractions are due to EC decay transitions to other excited states of the daughter nuclei. In the first case, the final state 7/2\,$^+$ with energy $E^{\,\prime}=93.1$~keV is the first excited state of the stable nucleus ${}^{107}_{\,\,\,47}$Ag with spin and parity 1/2\,$^-$; $\gamma$-transition $7/2\,^+\to 1/2\,^-$ between these states has half-life $T^{\,\gamma}_{1/2}=44.3$~s. In the second case, the mother nucleus ${}^{118m}_{\,\,\,\,\,\,\,\,51}$Sb is an isomeric state of the nucleus ${}^{118}_{\,\,\,51}$Sb (with spin and parity 1\,$^+$) with excitation energy $E^*=250$~keV, while the final state 7\,$^-$ of energy $E^{\,\prime}=2574.8$~keV is a highly excited state of stable nucleus ${}^{118}_{\,\,\,50}$Sn with spin and parity 0\,$^+$.  A $\gamma$-transition cascade  $7\,^-\to 4\,^+\to 2\,^+\to 0\,^+$ occurs in $\tau\sim 10^{-7}$~s.

\begin{table}
\caption{\label{t1} The pairs of mother and daughter nuclei, related by Gamow--Teller transition following the electron capture. Here $^A_ZX$ is the mother nucleus (of spin and parity $I^{\pi\strut}$ and half-life $T_{1/2}$), undergoing the Gamow--Teller transition into one state of the daughter nucleus $\lefteqn{{}_{Z-1}}\,\,\,\,\,{}^AX^{\prime}$ (of spin and parity $I^{\prime\pi}$ and of excitation energy $E^{\,\prime}$) with probability $P\ge 98~\%$. $Q_{EC}$ is the energy release in the transition. The data are taken from \cite{IAEA}.}
\begin{center}
\begin{tabular}{|r|r|r|r|r|r|r|r|}
\hline
${}^A_ZX$ & $I^{\pi\strut}$ & $T_{1/2}$ & $\lefteqn{{}_{Z-1}}\,\,\,\,\,{}^AX^{\prime}$ & $I^{\prime\pi}$ & $E^{\,\prime}$, keV & $Q_{EC}$, keV & $P$, \% \\
\hline
${}^{71}_{32}$Ge & $1/2^{\,-\strut}$ & 11.4~d & ${}^{71}_{31}$Ga & 3/2\,$^-$ & 0 & 232.6 & 100 \\
\hline
${}^{107}_{\,\,\,48}$Cd & $5/2^{\,+\strut}$ & 6.5~h & ${}^{107}_{\,\,\,47}$Ag$^*$ & 7/2\,$^+$ & 93.1 & 1323.2 & 99.7 \\
\hline
${}^{118m}_{\,\,\,\,\,\,\,\,51}$Sb & $8^{\,-\strut}$ & 5.0~h & ${}^{118}_{\,\,\,50}$Sn$^*$ & 7\,$^-$ & 2574.8 & 1332 & 98.3 \\
\hline
${}^{131}_{\,\,\,55}$Cs & $5/2^{\,+\strut}$ & 9.7~d & ${}^{131}_{\,\,\,54}$Xe & 3/2\,$^+$ & 0 & 354.8 & 100 \\
\hline
${}^{135}_{\,\,\,57}$La & $5/2^{\,+\strut}$ & 19.5~h & ${}^{135}_{\,\,\,56}$Ba & 3/2\,$^+$ & 0 & 1207 & 98.1 \\
\hline
${}^{163}_{\,\,\,68}$Er & $5/2^{\,-\strut}$ & 75~m & ${}^{163}_{\,\,\,67}$Ho & 7/2\,$^-$ & 0 & 1211 & 99.9 \\
\hline
${}^{165}_{\,\,\,68}$Er & $5/2^{\,-\strut}$ & 10.4~h & ${}^{165}_{\,\,\,67}$Ho & 7/2\,$^-$ & 0 & 378 & 100 \\
\hline
\end{tabular}
\end{center}
\end{table}

\section{Hyperfine structure and modulation \label{HFS}}

In this section we discuss the hyperfine structure of the ground state of a H-like ion and its effect on neutrino modulation possibilities. In the point-like nucleus approximation, the distance between two hyperfine levels is determined by the squared modulus $|\psi_{g.s.}(0)|^2$ of the electron wavefunction in the ion, taken at the origin (at the nucleus), and by the nuclear spin $I$ and magnetic moment $\mu$. The result is (see, e.g., \cite{Shabaev_1997})
\begin{equation}
\label{6}
\Delta_{HF}\equiv E_{F=I+1/2}-E_{F=I-1/2}=
\frac{4\alpha^4Z^3}{3}\,\frac{\mu}{\mu_N}\,\frac{m_e}{m_p}\,\frac{2I+1}{2I}\,m_ec^2\,A(\alpha Z),
\end{equation}
where $\alpha=e^2/(\hbar c)\simeq 1/137.036$ is the fine-structure constant, $Z$ is the nuclear charge, $\mu_N$ is the nuclear magneton, $m_p$ is the proton mass, $A(\alpha Z)$ is the relativistic factor, defined as
\begin{equation}
\label{7}
A(\alpha Z) = \frac{1}{\left(2\sqrt{1-\alpha^2 Z^2}-1\right)\sqrt{1-\alpha^2 Z^2}}\,.
\end{equation}
For ions with high $Z$, there are significant contributions (of the order of 5\%) into the hyperfine splitting from finite-size nuclear corrections and radiative corrections.

Note that the sign of the nuclear magnetic moment $\mu$ determines the sequence of hyperfine levels in a H-like ion: when $\mu>0$, the state with angular momentum $F=I-1/2$ is lower, while in the case of $\mu<0$ the lower state has $F=I+1/2$ (see Fig.~\ref{f1}). In Gamow--Teller transitions in H-like ions, EC decay from the lower state could be either allowed or forbidden. Accordingly, we define two types of H-like ions: A type and F type, depending on whether the EC decay is allowed or forbidden for the lower state.

The type of an ion is determined, first, by the sign of the nuclear magnetic moment $\mu$ and, second, by the relation between the initial and the final nuclear spins $I$ and $I^{\prime}$. All possible combinations are represented on Fig.~\ref{f1}. In these terms, only ions of A type (with $^{140}$Pr, $^{142}$Pm and $^{122}$I nuclei) were studied in \cite{Litvinov_2007,Litvinov_2008,Winckler_2009,Kienle_2013,Atanasov_2012}.

The modulation methods for neutrino beams, produced by H-like ions of A and F types, are different. Indeed, for A-type ions, the transition from the lower to the upper state, induced by a resonant electromagnetic field, is accompanied by an interruption of the neutrino emission. For F-type ions, on the contrary, such transition will initiate the neutrino emission. In both cases, the reverse transition from the upper level to the lower one could be either spontaneous or induced by an external resonant electromagnetic wave.

To our opinion, H-like ions of F type are of special interest. Such ions decay predominantly or exclusively from the upper state of the hyperfine structure. Hence, their lower state is practically stable. This circumstance may be used to accumulate ions in a storage ring. Then, by electromagnetic transitions to the upper state, one may form neutrino pulses.

The rate of a spontaneous $M1$ transition from the upper hyperfine state $|F_2f_2\rangle$ (with arbitrary projection $f_2$) to the lower state $|F_1f_1\rangle$ is (see, e.g., \cite{LL4})
\begin{equation}
\label{8}
w_{HF}(F_2\to F_1)=\frac{4\alpha|\Delta_{HF}|^3}{3\hbar (m_ec^2)^2}\cdot\frac{1}{2I+1}\cdot
\left\{\begin{array}{ll}
I, & \mbox{for}\quad F_2=I+1/2\to F_1=I-1/2,\\
I+1, & \mbox{for}\quad F_2=I-1/2\to F_1=I+1/2.
\end{array}\right.
\end{equation}
The resonant wavelength is $\lambda_{HF}=2\pi\hbar c/|\Delta_{HF}|$.

Let the electromagnetic wave, which induces the hyperfine transitions, be a right-polarized wave (propagating along $z$ axis) with the magnetic field
\begin{equation}
\label{9}
{\bf h}(t)=h_0\left({\bf e}_x\cos\omega t+{\bf e}_y\sin\omega t\right).
\end{equation}
Neglecting the nuclear magnetic moment $\mu$ compared to the electron magnetic moment $\mu_e=-\mu_B$ ($\mu_B$ is the Bohr magneton), we write down the interaction operator for the H-like ion and the electromagnetic wave as follows:
\begin{equation}
\label{10}
\hat V(t)=-\mu_e\mbox{\boldmath $\sigma$}{\bf h}(t),
\end{equation}
where $\sigma_i$ are the Pauli matrices. The state vector of the ion could be presented as a linear combination of the ion stationary states:
\begin{equation}
\label{11}
|\Psi(t)\rangle=\sum_{Ff}a_{Ff}(t)\,|Ff\rangle\, e^{-i\frac{\scriptstyle \mathstrut E_Ft}{\scriptstyle \mathstrut \hbar}}\,,
\end{equation}
with amplitudes $a_{Ff}(t)$, which describe transitions between the hyperfine states. The amplitudes obey the equation
\begin{equation}
\label{12}
{\dot a}_{Ff}(t)=-\frac{i}{\hbar}\sum_{F^{\prime}f^{\prime}}a_{F^{\prime}f^{\prime}}(t)
\langle Ff|{\hat V}(t)|F^{\prime}f^{\prime}\rangle\,
e^{i\frac{\scriptstyle \mathstrut (E_F-E_{F^{\prime}})t}{\scriptstyle \mathstrut\hbar}}\,.
\end{equation}

In the resonant case,
\begin{equation}
\label{13}
\omega=\frac{E_2-E_1}{\hbar}\equiv\frac{|\Delta_{HF}|}{\hbar},
\end{equation}
where $E_1$ and $E_2$ are the energies of the lower and upper hyperfine levels with angular momenta $F_1$ and $F_2$, respectively, the transitions between pairs of states $|F_1 f\rangle$ and $|F_2\,f+1\rangle$ dominate in the equations for the amplitudes. Right-polarized radiation initiates Rabi oscillations between these states, governed by the following equations:
\begin{equation}
\label{14}
\left\{\begin{array}{l}
{\dot a}_{F_2 f+1}(t)=-i\,\Omega^f_{F_1F_2}\,a_{F_1 f}(t),\\
{\dot a}_{F_1 f}(t)=-i\,\Omega^f_{F_1F_2}\,a_{F_2 f+1}(t).
\end{array}\right.
\end{equation}
The Rabi frequency takes the form
\begin{equation}
\label{15}
\Omega^f_{F_1F_2}=-\frac{\mu_Bh_0}{\hbar}\,
\sqrt{3(2F_1+1)}\,\,W(I\frac{1}{2}F_21,F_1\frac{1}{2})\,C^{F_2f+1}_{F_1f\,\,11}\,,
\end{equation}
where $W(abcd,ef)$ is the Racah function.

If at the time $t=0$ an ion is in the lower state $|F_1 f\rangle$, then, after applying a right-polarized field pulse of duration $T^f_{F_1F_2}/2=\pi/|\Omega^f_{F_1F_2}|$ ($\pi$-pulse), the ion transits to the upper state $|F_2\,f+1\rangle$, absorbing a photon with the energy $\hbar\Omega^f_{F_1F_2}$. The similar $\pi$-pulse will result in a transition from the upper state to the lower one, with a ''returning'' of a photon into the falling wave.

Equations (\ref{14}), (\ref{15}) imply that, with fixed $F_1$ and $F_2$ for any H-like ion, the oscillation period $T^f_{F_1F_2}$ depends on the projection $f$ of the lower hyperfine state $|F_1f\rangle$, which oscillates in coupling with the upper state $|F_2\, f+1\rangle$. There is, however, a simple case, when $F_2=F_1+1$ (this corresponds to $\mu>0$), and all the unexcited ions are in  the state with the maximal projection value $f=F_1$ (e.g. as the result of the use of optical pumping). Then the right-polarized radiation will cause the oscillations between the states $|F_1F_1\rangle$ and $|F_2F_2\rangle$ with the frequency
\begin{equation}
\label{16}
\Omega^{F_1}_{F_1F_2}=\frac{\mu_Bh_0}{\hbar}\,\sqrt{\frac{2I}{2I+1}}\,.
\end{equation}

A typical Rabi frequency $\Omega_0=\mu_Bh_0/\hbar$ corresponds to a period $T_0=2\pi\hbar/\mu_Bh_0$. Using the estimate for the radiation flux density $S=ch_0^2/(4\pi)\simeq 1$~W/cm$^2$, we get the characteristic oscillation period of $T_0\simeq 1.1\cdot 10^{-5}$~s. 

\section{Results and discussion \label{Results}}

Let us now discuss the properties of the selected H-like ions (see Table~\ref{t1}), which could be used as sources of monochromatic neutrino beams. Table~\ref{t2} gives magnetic moments $\mu$ of the mother nuclei (the data are taken from \cite{IAEA}); the sign of nuclear magnetic moment for $^{118m}$Sb isotope is unknown. In line with Fig.~\ref{f1}, we determine the type (shown in Table~\ref{t2}) for selected H-like ions (excluding $^{118m}$Sb), A or F, depending on the sign of magnetic moment and the relation between the initial and final nuclear spins $I$ and $I^{\prime}$.

\begin{table}
\caption{\label{t2} The main properties of the H-like ions --- possible sources of monochromatic and modulated neutrino beams. Here $^A_ZX$ is the mother nucleus, which undergoes the Gamow--Teller transition $I^{\pi\strut}\to\, I^{\prime\pi}$ into the final state of the daughter nucleus, $\mu$ is the magnetic moment of the mother nucleus in units of nuclear magneton $\mu_N$, A or F is the type of the H-like ion, $|\Delta_{HF}|$ is the absolute value of hyperfine splitting of the H-like ion ground state, $\lambda_{HF}$ is the resonant wavelength of hyperfine transition, $\tau_{HF}$ is the lifetime of the upper hyperfine state with respect to spontaneous emission.}
\begin{center}
\begin{tabular}{|r|c|r|c|c|c|c|}
\hline
${}^A_ZX$ & $I^{\pi\strut}\to\, I^{\prime\pi}$ & $\mu/\mu_N$ & Type & $|\Delta_{HF}|$, eV & $\lambda_{HF}$, $\mu$m & $\tau_{HF}$, s \\
\hline
${}^{71}_{32}$Ge & $1/2^{\,-\strut}\!\!\to\, 3/2^{\,-}$ & $+0.55$ & F & 0.041 & 30.2 & 1024\\
\hline
${}^{107}_{\,\,\,48}$Cd & $5/2^{\,+\strut}\!\!\to\,7/2^{\,+}$ & $-0.615$ & A & 0.105 & 11.8 & 26.3\\
\hline
${}^{118m}_{\,\,\,\,\,\,\,\,51}$Sb & $8^{\,-\strut}\!\!\to\,7^{\,-}$ & $2.32$ & & 0.433 & 2.86 & $0.46(+)$, $0.41(-)$ \\
\hline
${}^{131}_{\,\,\,55}$Cs & $5/2^{\,+\strut}\!\!\to\,3/2^{\,+}$ & $+3.54$ & A & 0.973 & 1.27 & 0.046\\
\hline
${}^{135}_{\,\,\,57}$La & $5/2^{\,+\strut}\!\!\to\,3/2^{\,+}$ & $+3.70$ & A & 1.162 & 1.06 & 0.027\\
\hline
${}^{163}_{\,\,\,68}$Er & $5/2^{\,-\strut}\!\!\to\,7/2^{\,-}$ & $+0.56$ & F & 0.346 & 3.58 &1.03 \\
\hline
${}^{165}_{\,\,\,68}$Er & $5/2^{\,-\strut}\!\!\to\,7/2^{\,-}$ & $+0.64$ & F & 0.399 & 3.10 & 0.67 \\
\hline
\end{tabular}
\end{center}
\end{table}

In Table~\ref{t2} the following ion parameters are also given: the hyperfine splitting $|\Delta_{HF}|$ of the ground state (calculated according to the Eqs. (\ref{6}) and (\ref{7})), the resonant wavelength $\lambda_{HF}$, the lifetime $\tau_{HF}=1/w_{HF}$ (see Eq.~(\ref{8})) of the upper hyperfine state with respect to spontaneous emission. For the ion with the nucleus $^{118m}$Sb we give values of $\tau_{HS}$ both for positive $(+)$ and for negative $(-)$ magnetic moment $\mu$.

It should be emphasized that all the given values are estimates, since the Eq.~(\ref{6}) for the hyperfine splitting $\Delta_{HF}$ (which is used to calculate the wavelength $\lambda_{HF}$ and the transition rate $w_{HF}$) does not take into account the nuclear and radiative corrections. We also ignore statistical and systematical uncertaincies in the measured values of nuclear magnetic moments.

All the selected nuclei have rather large half-lives (for a nucleus, accelerated to Lorentz factor $\gamma$, lifetime increases by $\gamma$ times). This means that the total number of radioactive ions in a storage ring should be large, in order to get the neutrino beam of significant intensity. Of course, the required beam intensity, as well as its energy (and, thus, the factor $\gamma$), depends on the experimental problem.

Note also that the values $T_{1/2}$ in Table~\ref{t1} correspond to the nuclei in neutral atoms, while the half-lives of the nuclei in H-like ions of A type is shorter by the factor of the order of 1.5 (see Refs.~\cite{Litvinov_2007,Patyk_2008,Ivanov_2008,Siegien_2011}).

To demonstrate the possibility of using the selected isotopes as neutrino sources, let us consider the following example. In Refs. \cite{Sato_2005,Rolinec_2007}, the authors studied the possibility of measuring the neutrino oscillation parameters, using a fully-monochromatic neutrino beam ($E^0_{\nu}=267$~keV), which is produced by EC decay of accelerated ions of  $^{110}$Sn (with the nucleus of zero spin) with the half-life $T_{1/2}=4.1$~h (this is comparable to the half-lives of the nuclei, selected in this work). The ions were supposed to be accelerated to Lorentz factors $\gamma\ge 1000$. The conclusion is that such measurement is possible, though it is difficult to conduct (in the subsequent literature on this subject, the interest shifted from monochromatic EC-beams to non-monochromatic $\beta$-beams, which could be made much more intense with the use of short-living isotopes; see, e.g., \cite{Estevez_2011}).

Of all the isotopes selected here, $^{163}$Er has the minimal half-life $T_{1/2}=75$~m. The H-like ion of this nucleus is of F type, i.e. its lower hyperfine state is almost stable with respect to EC decay. Therefore, such ions could be accumulated in a storage ring before they are used to produce modulated monochromatic neutrino radiation. The considered ion satisfies the condition $F_2=F_1+1$, where $F_2=3$ and $F_1=2$ are the total angular momenta of the upper and lower hyperfine states. Hence, one can obtain Rabi oscillations between only two hyperfine states $|F_1F_1\rangle$ and $|F_2F_2\rangle$, using right-polarized electromagnetic radiation. The transfer of all ions into the EC-unstable upper hyperfine state by means of a $\pi$-pulse will cause neutrino emission for a time period of $\tau_{HS}\simeq 1$~s (the mean time of spontaneous transition). Thus, in this scheme, neutrinos are produced in pulses of duration $\tau_{HS}\simeq 1$~s.

For the A-type ions, the situation seems to be more complicated. Among them, the ion of $^{107}$Cd has the shortest lifetime. However, the lifetime of the upper hyperfine state $\tau_{HS}=26.3$~s is rather large. So in this case one should modulate the neutrino beam by $\pi$-pulses, transfering ions from the lower states to the upper ones and vice versa. Unfortunately, the nuclear magnetic moment of $^{107}$Cd is negative, so that the condition $F_2=F_1+1$ is not satisfied. This somewhat reduces the efficiency of modulation.

The nucleus $^{135}$La has a larger half-life, than $^{107}$Cd, and the positive magnetic moment. However, the lifetime of the upper hyperfine state is only $\tau_{HS}=0.027$~s. This will be the time of neutrino beam interruption, after transferring of all ions to the upper hyperfine state by a $\pi$-pulse. As for the $^{71}$Ge nucleus, with the positive magnetic moment and the spontaneous emission time $\tau_{HS}=1024$~s, its large half-life will hardly allow obtaining a neutrino beam of high intensity.

\section{Conclusion \label{Conclusion}}

The use of radioactive ions leads to possibility of producing clean neutrino beams with well-known spectra. Such beams may be useful for a wide range of neutrino researches. Particular attention should be paid to monochromatic and modulated neutrino beams. As a source of such beams, one can use H-like ions with electron-capturing nuclei, which decay through Gamow--Teller transitions. Electromagnetic modulation is possible due to the features of the hyperfine structure of H-like ions. 

The nuclei of such ions have to satisfy several requirements. The mother nuclei must have non-zero spin. The spins of the initial and the final nuclear states have to differ by~1, their parities have to be the same. To provide high monochromaticity of the neutrino beam, $\beta^+$ decay must be highly suppressed and the nuclear transition through electron capture must go predominantly to only one final state (we accepted the branching larger than 98\% of total decay probability). In order to produce a beam of significant intensity, the nuclei have to be rather short-living (we set the upper limit on the half-lives \mbox{$T_{1/2}< 10^6~\mbox{s}\simeq 11.6$~d}).

In this paper, we have found all the nuclei, satisfying the stated requirements. Using H-like ions with these nuclei, one can obtain almost monochromatic neutrino beams that could be modulated with electromagnetic radiation, by inducing the transitions between the hyperfine levels in the ions. Modulation of a neutrino beam may significantly increase the efficiency of event registration in neutrino experiments by suppressing constant and random background noise.

It turns out that the list of appropriate nuclei is rather short. However, the H-like ions with these nuclei noticeably differ from each other both by the lifetime with respect to the electron capture as well as by the hyperfine structure of the ground states. The latter is important for the modulation method. Given a specific problem, one can choose a suitable sort of H-like ions from our list and use it as a source of modulated monochromatic neutrino beams.

\section*{Acknowledgements}

We are grateful to I.M.~Pavlichenkov, who called our attention to this problem, and to M.D.~Skorokhvatov for his interest to our research. The work was partially supported by the grant NSh-932.2014.2 from MES of Russia.

\end{document}